\documentclass{article}
\usepackage{smc2017}
\usepackage{times}
\usepackage{ifpdf}
\usepackage[english]{babel}
\usepackage{cite}
\usepackage{booktabs,hhline,tabularx,multirow}

\def\papertitle{SAMPLE-LEVEL DEEP CONVOLUTIONAL NEURAL NETWORKS FOR MUSIC AUTO-TAGGING USING RAW WAVEFORMS}
\def\firstauthor{Jongpil Lee}
\def\secondauthor{Jiyoung Park}
\def\thirdauthor{Keunhyoung Luke Kim}
\def\fourthauthor{Juhan Nam}

\newif\ifpdf
\ifx\pdfoutput\relax
\else
   \ifcase\pdfoutput
      \pdffalse
   \else
      \pdftrue
\fi

\ifpdf 
  \usepackage[pdftex,
    pdftitle={\papertitle},
    pdfauthor={\firstauthor, \secondauthor, \thirdauthor},
    bookmarksnumbered, 
    pdfstartview=XYZ 
   ]{hyperref}

  \usepackage[pdftex]{graphicx}
  \graphicspath{{./figures/}}
  \DeclareGraphicsExtensions{.pdf,.jpeg,.png}

  \usepackage[figure,table]{hypcap}

\else 
  \usepackage[dvips,
    bookmarksnumbered, 
    pdfstartview=XYZ 
  ]{hyperref}  

  \usepackage[dvips]{epsfig,graphicx}
  \graphicspath{{./figures/}}
  \DeclareGraphicsExtensions{.eps}

  \usepackage[figure,table]{hypcap}
\fi

\hypersetup{
    colorlinks,%
    citecolor=black,%
    filecolor=black,%
    linkcolor=black,%
    urlcolor=black
}

\title{\papertitle}

\oneauthor
   {\firstauthor \hspace{1cm} \secondauthor \hspace{1cm} \thirdauthor \hspace{1cm} \fourthauthor} {Korea Advanced Institute of Science and Technology (KAIST) \\ %
     {\tt {[richter, jypark527, dilu, juhannam]@kaist.ac.kr}}}

\begin{document}
\capstartfalse
\maketitle
\capstarttrue
\begin{abstract}

Recently, the end-to-end approach that learns hierarchical representations from raw data using deep convolutional neural networks has been successfully explored in the image, text and speech domains. This approach was applied to musical signals as well but has been not fully explored yet. To this end, we propose sample-level deep convolutional neural networks which learn representations from very small grains of waveforms (e.g. 2 or 3 samples) beyond typical frame-level input representations. Our experiments show how deep architectures with sample-level filters improve the accuracy in music auto-tagging and they provide results comparable to previous state-of-the-art performances for the Magnatagatune dataset and Million Song Dataset. In addition, we visualize filters learned in a sample-level DCNN in each layer to identify hierarchically learned features and show that they are sensitive to log-scaled frequency along layer, such as mel-frequency spectrogram that is widely used in music classification systems.

\end{abstract}

\section{Introduction}\label{sec:introduction}
In music information retrieval (MIR) tasks, raw waveforms of music signals are generally converted to a time-frequency representation and used as input to the system. The majority of MIR systems use a log-scaled representation in frequency such as mel-spectrograms and constant-Q transforms and then compress the amplitude with a log scale. The time-frequency representations are often  transformed further into more compact forms of audio features depending on the task. All of these processes are designed based on acoustic knowledge or engineering efforts.  

Recent advances in deep learning, especially the development of deep convolutional neural networks (DCNN), made it possible to learn the entire hierarchical representations from the raw input data, thereby minimizing the input data processing by hands. This end-to-end hierarchical learning was attempted early in the image domain, particularly since the DCNN achieves break-through results in image classification \cite{krizhevsky2012imagenet}. These days, the method of stacking small filters (e.g. 3x3) is widely used after it has been found to be effective in learning more complex hierarchical filters while conserving receptive fields\cite{simonyan2014very}. In the text domain, the language model typically consists of two steps: word embedding and word-level learning. While word embedding plays a very important role in language processing \cite{mikolov2013distributed}, it has limitations in that it is learned independently from the system. Recent work using CNNs that take character-level text as input showed that the end-to-end learning approach can yield comparable results to the word-level learning system \cite{zhang2015character,kim2015character}. In the audio domain, learning from raw audio has been explored mainly in the automatic speech recognition task \cite{palaz2015convolutional,palaz2015analysis,collobert2016wav2letter,van2016wavenet,sainath2015learning}. They reported that the performance can be similar to or even superior to that of the models using spectral-based features as input. 

This end-to-end learning approach has been applied to music classification tasks as well\cite{dieleman2014end,ardilaaudio}. In particular, Dieleman and Schrauwen used raw waveforms as input of CNN models for music auto-tagging task and attempted to achieve comparable results to those using mel-spectrograms as input\cite{dieleman2014end}. Unfortunately, they failed to do so and attributed the result to three reasons. First, their CNN models were not sufficiently expressive (e.g. a small number of layers and filters) to learn the complex structure of polyphonic music. Second, they could not find an appropriate non-linearity function that can replace the log-based amplitude compression in the spectrogram. Third, the bottom layer in the networks takes raw waveforms in frame-level which are typically several hundred samples long. The filters in the bottom layer should learn all possible phase variations of periodic waveforms which are likely to be prevalent in musical signals. The phase variations within a frame (i.e. time shift of periodic waveforms) are actually removed in the spectrogram.  

In this paper, we address these issues with sample-level DCNN. What we mean by ``sample-level'' is that the filter size in the bottom layer may go down to several samples long. We assume that this small granularity is analogous to pixel-level in image or character-level in text. We show the effectiveness of the sample-level DCNN in music auto-tagging task by decreasing strides of the first convolutional layer from frame-level to sample-level and accordingly increasing the depth of layers. Our experiments show that the depth of architecture with sample-level filters is proportional to the accuracy and also the architecture achieves results comparable to previous state-of-the-art performances for the MagnaTagATune dataset and the Million Song Dataset. In addition, we visualize filters learned in the sample-level DCNN.

\begin{figure*}[t]
\centering
\includegraphics[width=2.1\columnwidth]{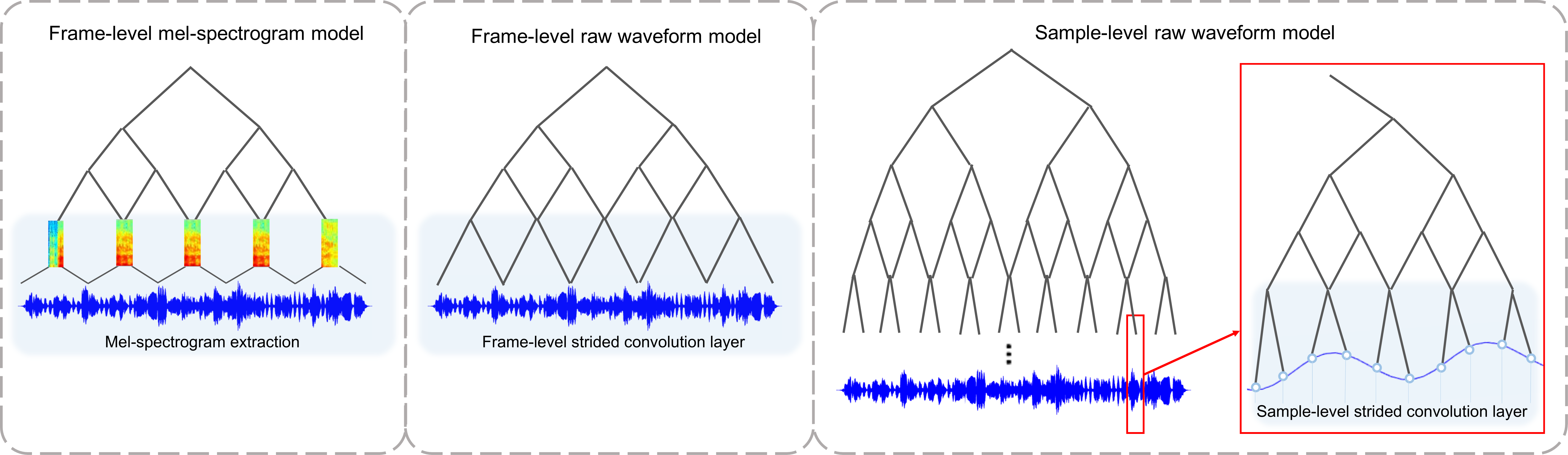}
\caption{Simplified model comparison of frame-level approach using mel-spectrogram (left), frame-level approach using raw waveforms (middle) and sample-level approach using raw waveforms (right).
\label{fig:fig1}}
\end{figure*}

\section{Related work}\label{sec:relatedwork}

Since audio waveforms are one-dimensional data, previous work that takes waveforms as input used a CNN that consists of one-dimensional convolution and pooling stages. While the convolution operation and filter length in upper layers are usually similar to those used in the image domain, the bottom layer that takes waveform directly conducted a special operation called \emph{strided convolution}, which takes a large filter length and strides it as much as the filter length (or the half). This frame-level approach is comparable to hopping windows with 100\% or 50\% hop size in a short-time Fourier transform. In many previous works, the stride and filter length of the first convolution layer was set to 10-20 ms (160-320 samples at 16 kHz audio) \cite{dieleman2014end,ardilaaudio,sainath2015learning,collobert2016wav2letter}. 

In this paper, we reduce the filter length and stride of the first convolution layer to the sample-level, which can be as small as 2 samples. Accordingly, we increase the depth of layers in the CNN model. There are some works that use 0.6 ms (10 samples at 16 kHz audio) as a stride length \cite{palaz2015convolutional,palaz2015analysis}, but they used a CNN model only with three convolution layers, which is not sufficient to learn the complex structure of musical signals.

\section{Learning Models}\label{sec:model}

Figure \ref{fig:fig1} illustrates three CNN models in the music auto-tagging task we compare in our experiments. In this section, we describe the three models in detail.

\subsection{Frame-level mel-spectrogram model}
This is the most common CNN model used in music auto-tagging. Since the time-frequency representation is two dimensional data, previous work regarded it as either two-dimensional images  or one-dimensional sequence of vectors \cite{pons2016experimenting,dieleman2014end,choi2016automatic,choi2016convolutional}. We only used one-dimensional(1D) CNN model for experimental comparisons in our work because the performance gap between 1D and 2D models is not significant and 1D model can be directly compared to models using raw waveforms.


\subsection{Frame-level raw waveform model}

In the frame-level raw waveform model, a strided convolution layer is added beneath the bottom layer of the frame-level mel-spectrogram model. The strided convolution layer is expected to learn  a filter-bank representation that correspond to filter kernels in a time-frequency representation. In this model, once the raw waveforms pass through the first strided convolution layer, the output feature map has the same dimensions as the mel-spectrogram. This is because the stride, filter length, and number of filters of the first convolution layer correspond to the hop size, window size, and number of mel-bands in the mel-spectrogram, respectively. This configuration was used for music auto-tagging task in \cite{dieleman2014end,ardilaaudio} and so we used it as a baseline model.

\subsection{Sample-level raw waveform model}
As described in Section \ref{sec:introduction}, the approach using the raw waveforms should be able to address log-scale amplitude compression and phase-invariance. Simply adding a strided convolution layer is not sufficient to overcome the problems. To improve this, we add multiple layers beneath the frame-level such that the first convolution layer can handle much smaller length of samples. For example, if the stride of the first convolution layer is reduced from 729 ($=3^6$) to 243 ($=3^5$), 3-size convolution layer and max-pooling layer are added to keep the output dimensions in the subsequent convolution layers unchanged. If we repeatedly reduce the stride of the first convolution layer this way, six convolution layers (five pairs of 3-size convolution and max-pooling layer following one 3-size strided convolution layer) will be added (we assume that the temporal dimensionality reduction occurs only through max-pooling and striding while zero-padding is used in convolution to preserve the size). We describe more details on the configuration strategy of sample-level CNN model in the following section.


\subsection{Model Design}\label{sec:modeldesign}

Since the length of an audio clip is variable in general, the following issues should be considered when configuring the temporal CNN architecture:
\begin{itemize}
   \item Convolution filter length and sub-sampling length
   \item The temporal length of hidden layer activations on the last sub-sampling layer
   \item The segment length of audio that corresponds to the input size of the network
\end{itemize}

First, we attempted a very small filter length in convolutional layers and sub-sampling length, following the VGG net that uses filters of $3\times3$ size and max-pooling of $2\times2$ size \cite{simonyan2014very}. Since we use one-dimensional convolution and sub-sampling for raw waveforms, however, the filter length and pooling length need to be varied. We thus constructed several DCNN models with different filter length and pooling length from 2 to 5, and verified the effect on music auto-tagging performance. As a sub-sampling method, max-pooling is generally used. Although sub-sampling using strided convolution has recently been proposed in a generative model\cite{van2016wavenet}, our preliminary test showed that max-pooling was superior to the stride sub-sampling method. In addition, to avoid exhausting model search, a pair of single convolution layer and max-pooling layer with the same size was used as a basic building module of the DCNN.


Second, the temporal length of hidden layer activations on the last sub-sampling layer reflects the temporal compression of the input audio by successive sub-sampling. We set the CNN models such that the temporal length of hidden layer activation is one. By building the models this way, we can significantly reduce the number of parameters between the last sub-sampling layer and the output layer. Also, we can examine the performance only by the depth of layers and the stride of first convolution layer. 


Third, in music classification tasks, the input size of the network is an important parameter that determines the classification performance \cite{P.Hamel:11,lee2017multi}. In the mel-spectrogram model, one song is generally divided into small segments with 1 to 4 seconds. The segments are used as the input for training and the predictions over all segments in one song are averaged in testing. In the models that use raw waveform, the learning ability according to the segment size has been not reported yet and thus we need to examine different input sizes when we configure the CNN models.




Considering all of these issues, we construct $m^n$-DCNN models where $m$ refers to the filter length and pooling length of intermediate convolution layer modules and $n$ refers to the number of the modules (or depth). An example of $m^n$-DCNN models is shown in Table \ref{table:table1} where $m$ is 3 and $n$ is 9. According to the definition, the filter length and pooling length of the convolution layer are 3 other than the first strided convolution layer. If the hop size (stride length) of the first strided convolution layer is 3, the time-wise output dimension of the convolution layer becomes 19683 when the input of the network is 59049 samples. We call this ``$3^9$ model with 19683 frames and 59049 samples as input''.


\begin{table}[t]
\centering
\label{table:table1}
\begin{tabular}{@{}cccc@{}}
\cmidrule(){1-4}
\multicolumn{4}{c}{\textbf{\boldmath{$3^9$} model, 19683 frames}}                                                                                                                                            \\
\multicolumn{4}{c}{\textbf{59049 samples (2678 ms) as input}}                                                                                                                                             \\ \midrule
layer                                                           & stride                                        & output                                                       & \# of params \\ \cmidrule(){1-4}
conv 3-128                                                       & $3$                                             & $19683\times128$                                                    & $512$          \\ \midrule
\begin{tabular}[c]{@{}c@{}}conv 3-128\\ maxpool 3\end{tabular}   & \begin{tabular}[c]{@{}c@{}}$1$\\ $3$\end{tabular} & \begin{tabular}[c]{@{}c@{}}$19683\times128$\\ $6561\times128$\end{tabular} & $49280$        \\ \midrule
\begin{tabular}[c]{@{}c@{}}conv 3-128\\ maxpool 3\end{tabular}   & \begin{tabular}[c]{@{}c@{}}$1$\\ $3$\end{tabular} & \begin{tabular}[c]{@{}c@{}}$6561\times128$\\ $2187\times128$\end{tabular}  & $49280$        \\ \midrule
\begin{tabular}[c]{@{}c@{}}conv 3-256\\ maxpool 3\end{tabular}   & \begin{tabular}[c]{@{}c@{}}$1$\\ $3$\end{tabular} & \begin{tabular}[c]{@{}c@{}}$2187\times256$\\ $729\times256$\end{tabular}   & $98560$        \\ \midrule
\begin{tabular}[c]{@{}c@{}}conv 3-256\\ maxpool 3\end{tabular}   & \begin{tabular}[c]{@{}c@{}}$1$\\ $3$\end{tabular} & \begin{tabular}[c]{@{}c@{}}$729\times256$\\ $243\times256$\end{tabular}    & $196864$       \\ \midrule
\begin{tabular}[c]{@{}c@{}}conv 3-256\\ maxpool 3\end{tabular}   & \begin{tabular}[c]{@{}c@{}}$1$\\ $3$\end{tabular} & \begin{tabular}[c]{@{}c@{}}$243\times256$\\ $81\times256$\end{tabular}     & $196864$       \\ \midrule
\begin{tabular}[c]{@{}c@{}}conv 3-256\\ maxpool 3\end{tabular}   & \begin{tabular}[c]{@{}c@{}}$1$\\ $3$\end{tabular} & \begin{tabular}[c]{@{}c@{}}$81\times256$\\ $27\times256$\end{tabular}      & $196864$       \\ \midrule
\begin{tabular}[c]{@{}c@{}}conv 3-256\\ maxpool 3\end{tabular}   & \begin{tabular}[c]{@{}c@{}}$1$\\ $3$\end{tabular} & \begin{tabular}[c]{@{}c@{}}$27\times256$\\ $9\times256$\end{tabular}       & $196864$       \\ \midrule
\begin{tabular}[c]{@{}c@{}}conv 3-256\\ maxpool 3\end{tabular}   & \begin{tabular}[c]{@{}c@{}}$1$\\ $3$\end{tabular} & \begin{tabular}[c]{@{}c@{}}$9\times256$\\ $3\times256$\end{tabular}        & $196864$       \\ \midrule
\begin{tabular}[c]{@{}c@{}}conv 3-512\\ maxpool 3\end{tabular}   & \begin{tabular}[c]{@{}c@{}}$1$\\ $3$\end{tabular} & \begin{tabular}[c]{@{}c@{}}$3\times512$\\ $1\times512$\end{tabular}        & $393728$       \\ \midrule
\begin{tabular}[c]{@{}c@{}}conv 1-512\\ dropout 0.5\end{tabular} & \begin{tabular}[c]{@{}c@{}}$1$\\ $-$\end{tabular} & \begin{tabular}[c]{@{}c@{}}$1\times512$\\ $1\times512$\end{tabular}        & $262656$        \\ \midrule
sigmoid                                                   & $-$                                             & $50$                                                           & $25650$        \\ \cmidrule(){1-4}\morecmidrules\cmidrule(){1-4}
\multicolumn{3}{c}{Total params}                                                                                                                                               & $1.9\times10^6$   \\ \cmidrule(){1-4}\morecmidrules\cmidrule(){1-4}
\end{tabular}
\caption{Sample-level CNN configuration. For example, in the layer column, the first 3 of ``conv 3-128'' is the filter length, 128 is the number of filters, and 3 of ``maxpool 3'' is the pooling length.
}
\end{table}

\begin{table*}[!t]
\centering
\label{table:table2}
\resizebox{\textwidth}{!}{\begin{tabular}{@{}cccccccccc@{}}
\toprule
\multicolumn{10}{c}{\textbf{\boldmath{$2^n$} models}}                                                                                 \\ \midrule
\multicolumn{5}{c}{model with 16384 samples (743 ms) as input} & \multicolumn{5}{c}{model with 32768 samples (1486 ms) as input} \\ \cmidrule(r){1-5} \cmidrule(l){6-10}
model     & $n$        & layer       & filter length \& stride      & AUC         & model       & $n$        & layer       & filter length \& stride      & AUC         \\ \cmidrule(r){1-5} \cmidrule(l){6-10} \morecmidrules\cmidrule(r){1-5} \cmidrule(l){6-10}
64 frames   & 6       & 1+6+1       & 256           & 0.8839      & 128 frames     & 7     & 1+7+1       & 256           &   0.8834      \\
128 frames  & 7       & 1+7+1       & 128           & 0.8899      & 256 frames     & 8     & 1+8+1       & 128           &    0.8872     \\
256 frames  & 8       & 1+8+1       & 64            & 0.8968      & 512 frames     & 9     & 1+9+1       & 64            &   0.8980       \\
512 frames  & 9       & 1+9+1       & 32            & 0.8994      & 1024 frames    & 10     & 1+10+1      & 32            &   0.8988       \\
1024 frames & 10       & 1+10+1      & 16            & 0.9011      & 2048 frames    & 11     & 1+11+1      & 16            &  0.9017    \\
2048 frames  & 11      & 1+11+1      & 8             & 0.9031      & 4096 frames    & 12     & 1+12+1      & 8             &  0.9031    \\
4096 frames & 12       & 1+12+1      & 4             & 0.9036      & 8192 frames    & 13     & 1+13+1      & 4       &  0.9039\\ 
8192 frames & 13       & 1+13+1      & 2             & 0.9032      & 16384 frames    & 14     & 1+14+1      & 2      &  0.9040 \\

\toprule
\multicolumn{10}{c}{\textbf{\boldmath{$3^n$} models}}                                                                                 \\ \midrule
\multicolumn{5}{c}{model with 19683 samples (893 ms) as input} & \multicolumn{5}{c}{model with 59049 samples (2678 ms) as input} \\ \cmidrule(r){1-5} \cmidrule(l){6-10}
model       & $n$       & layer       & filter length \& stride      & AUC         & model      & $n$         & layer       & filter length \& stride      & AUC         \\ \cmidrule(r){1-5} \cmidrule(l){6-10} \morecmidrules\cmidrule(r){1-5} \cmidrule(l){6-10}
27 frames   & 3       & 1+3+1       & 729           & 0.8655      & 81 frames      & 4    & 1+4+1       & 729           & 0.8655            \\
81 frames   & 4      & 1+4+1       & 243           & 0.8753      & 243 frames      & 5    & 1+5+1       & 243           & 0.8823            \\
243 frames  & 5       & 1+5+1       & 81            & 0.8961      & 729 frames     & 6     & 1+6+1       & 81            & 0.8936            \\
729 frames  & 6       & 1+6+1       & 27            & 0.9012      & 2187 frames    & 7     & 1+7+1      & 27            & 0.9002            \\
2187 frames & 7       & 1+7+1      & 9            & 0.9033      & 6561 frames      & 8   & 1+8+1      & 9            & 0.9030            \\
6561 frames & 8       & 1+8+1      & 3             & 0.9039      & 19683 frames    & 9     & 1+9+1      & 3             & \textbf{0.9055}            \\

\toprule
\multicolumn{10}{c}{\textbf{\boldmath{$4^n$} models}}                                                                                 \\ \midrule
\multicolumn{5}{c}{model with 16384 samples (743 ms) as input} & \multicolumn{5}{c}{model with 65536 samples (2972 ms) as input} \\ \cmidrule(r){1-5} \cmidrule(l){6-10}
model        & $n$      & layer       & filter length \& stride      & AUC         & model       & $n$        & layer       & filter length \& stride      & AUC         \\ \cmidrule(r){1-5} \cmidrule(l){6-10} \morecmidrules\cmidrule(r){1-5} \cmidrule(l){6-10}
64 frames    & 3      & 1+3+1       & 256           & 0.8828      & 256 frames     & 4     & 1+4+1       & 256           & 0.8813            \\
256 frames   & 4      & 1+4+1       & 64            & 0.8968      & 1024 frames    & 5      & 1+5+1       & 64            & 0.8950        \\
1024 frames  & 5      & 1+5+1      & 16            & 0.9010     & 4096 frames      & 6   & 1+6+1      & 16            &   0.9001          \\
4096 frames  & 6      & 1+6+1      & 4             & 0.9021      & 16384 frames    & 7     & 1+7+1      & 4             & 0.9026            \\

\toprule
\multicolumn{10}{c}{\textbf{\boldmath{$5^n$} models}}                                                                                 \\ \midrule
\multicolumn{5}{c}{model with 15625 samples (709 ms) as input} & \multicolumn{5}{c}{model with 78125 samples (3543 ms) as input} \\ \cmidrule(r){1-5} \cmidrule(l){6-10}
model        & $n$      & layer       & filter length \& stride      & AUC         & model     & $n$          & layer       & filter length \& stride      & AUC         \\ \cmidrule(r){1-5} \cmidrule(l){6-10} \morecmidrules\cmidrule(r){1-5} \cmidrule(l){6-10}
125 frames   & 3      & 1+3+1       & 125           & 0.8901      & 625 frames     & 4     & 1+4+1       & 125           & 0.8870            \\
625 frames   & 4      & 1+4+1       & 25            & 0.9005      & 3125 frames    & 5      & 1+5+1       & 25            & 0.9004            \\
3125 frames  & 5       & 1+5+1       & 5            & 0.9024      & 15625 frames   & 6      & 1+6+1      & 5            & 0.9041            \\
\bottomrule

\end{tabular}}
\caption{Comparison of various $m^n$-DCNN models with different input sizes. $m$ refers to the filter length and pooling length of intermediate convolution layer modules and $n$ refers to the number of the modules. Filter length \& stride indicates the value of the first convolution layer. In the layer column, the first digit '1' of 1+$n$+1 is the strided convolution layer, and the last digit '1' is convolution layer which actually works as a fully-connected layer.}
\end{table*}

\section{Experimental Setup}

In this section, we introduce the datasets used in our experiments and describe experimental settings.

\subsection{Datasets}
We evaluate the proposed model on two datasets, MagnaTagATune dataset (MTAT) \cite{law2009evaluation} and Million Song Dataset (MSD) annotated with the Last.FM tags \cite{bertin2011million}. We primarily examined the proposed model on MTAT and then verified the effectiveness of our model on MSD which is much larger than MTAT\footnote{MTAT contains 170 hours long audio and MSD contains 1955 hours long audio in total}. We filtered out the tags and used most frequently labeled 50 tags in both datasets, following the previous work \cite{dieleman2014end}, \cite{choi2016automatic,choi2016convolutional}\footnote{ \url{https://github.com/keunwoochoi/MSD_split_for_tagging}}. Also, all songs in the two datasets were trimmed to 29.1 second long and resampled to 22050 Hz as needed. We used AUC (Area Under Receiver Operating Characteristic) as a primary evaluation metric for music auto-tagging. 

\subsection{Optimization}
We used sigmoid activation for the output layer and binary cross entropy loss as the objective function to optimize. For every convolution layer, we used batch normalization \cite{ioffe2015batch} and ReLU activation. We should note that, in our experiments, batch normalization plays a vital role in training the deep models that takes raw waveforms. We applied dropout of 0.5 to the output of the last convolution layer and minimized the objective function using stochastic gradient descent with 0.9 Nesterov momentum. The learning rate was initially set to 0.01 and decreased by a factor of 5 when the validation loss did not decrease more than 3 epochs. A total decrease of 4 times, the learning rate of the last training was 0.000016. Also, we used batch size of 23 for MTAT and 50 for MSD, respectively. In the mel-spectrogram model, we conducted the input normalization simply by dividing the standard deviation after subtracting mean value of entire input data. On the other hand, we did not perform the input normalization for raw waveforms.

\section{Results}
In this section, we examine the proposed models and compare them to previous state-of-the-art results.

\begin{table}[t]
\centering
\resizebox{\columnwidth}{!}{\begin{tabular}{@{}ccccc@{}}
\toprule
\textbf{\boldmath{\begin{tabular}[c]{@{}c@{}}$3^n$ models,\\ 59049 samples\\ as input\end{tabular}}}      & \boldmath{$n$} & \textbf{\begin{tabular}[c]{@{}c@{}}window\\ (filter length)\end{tabular}} & \textbf{\begin{tabular}[c]{@{}c@{}}hop\\ (stride)\end{tabular}} & \textbf{AUC} \\ \midrule
\multirow{5}{*}{\begin{tabular}[c]{@{}c@{}}Frame-level\\ (mel-spectrogram)\end{tabular}} & 4   & 729                                                                       & 729                                                             & 0.9000       \\
                                                                                         & 5   & 729                                                                       & 243                                                             & 0.9005       \\
                                                                                         & 5   & 243                                                                       & 243                                                             & 0.9047       \\
                                                                                         & 6   & 243                                                                       & 81                                                              & \textbf{0.9059}       \\
                                                                                         & 6   & 81                                                                        & 81                                                              & 0.9025       \\ \midrule
\multirow{5}{*}{\begin{tabular}[c]{@{}c@{}}Frame-level\\   (raw waveforms)\end{tabular}}   & 4   & 729                                                                       & 729                                                             & 0.8655       \\
                                                                                         & 5   & 729                                                                       & 243                                                             & 0.8742       \\
                                                                                         & 5   & 243                                                                       & 243                                                             & 0.8823       \\
                                                                                         & 6   & 243                                                                       & 81                                                              & 0.8906       \\
                                                                                         & 6   & 81                                                                        & 81                                                              & 0.8936       \\ \midrule
\multirow{3}{*}{\begin{tabular}[c]{@{}c@{}}Sample-level\\ (raw waveforms)\end{tabular}}  
						& 7   & 27                                                                        & 27                                                              & 0.9002       \\
                                                                                         & 8   & 9                                                                         & 9                                                               & 0.9030       \\
                                                                                         & 9   & 3                                                                         & 3                                                               & \textbf{0.9055}       \\ \bottomrule
\end{tabular}}
\caption{Comparison of three CNN models with different window (filter length) and hop (stride) sizes. $n$ represents the number of intermediate convolution and max-pooling layer modules, thus $3 ^n$ times hop (stride) size of each model is equal to the number of input samples.}
\label{table:table3}
\end{table}

\begin{table}[t]
\centering
\resizebox{\columnwidth}{!}{\begin{tabular}{@{}cccc@{}}
\toprule
\textbf{input type}                                                                      & \textbf{model}        & \textbf{MTAT}    & \textbf{MSD}    \\ \midrule
\multirow{4}{*}{\begin{tabular}[c]{@{}c@{}}Frame-level\\ (mel-spectrogram)\end{tabular}} & Persistent CNN\cite{liu2016applying}        & 0.9013          & -           \\
                   & 2D CNN\cite{choi2016automatic}           & 0.894          & 0.851           \\
                                                                                       \vspace{1.6 mm}  &  CRNN\cite{choi2016convolutional}                 & -              &  0.862               \\
                                                                                       \vspace{1.6 mm}  & Proposed DCNN    & \textbf{0.9059}          & -               \\ \midrule
\begin{tabular}[c]{@{}c@{}}Frame-level\\ (raw waveforms)\end{tabular}                    & 1D CNN\cite{dieleman2014end} & 0.8487             & -               \\ \midrule
\begin{tabular}[c]{@{}c@{}}Sample-level\\ (raw waveforms)\end{tabular}                   & Proposed DCNN          & \textbf{0.9055} & \textbf{0.8812} \\ \bottomrule
\end{tabular}}
\caption{Comparison of our works to prior state-of-the-arts}
\label{table:table4}
\end{table}

\subsection{\boldmath{$m^n$}-DCNN models}
Table 2 shows the evaluation results for the $m^n$-DCNN models on MTAT for different input sizes, number of layers, filter length and stride of the first convolution layer. As described in Section \ref{sec:modeldesign}, $m$ refers to the filter length and pooling length of intermediate convolution layer modules and $n$ refers to the number of the modules. In Table 2, we can first find that the accuracy is proportional to $n$ for most $m$. Increasing $n$ given $m$ and input size indicates that the filter length and stride of the first convolution layer become closer to the sample-level (e.g. 2 or 3 size). When the first layer reaches the small granularity, the architecture is seen as a model constructed with the same filter length and sub-sampling length in all convolution layers as depicted in Table \ref{table:table1}. The best results were obtained when $m$ was 3 and $n$ was 9. Interestingly, the length of 3 corresponds to the 3-size spatial filters in the VGG net\cite{simonyan2014very}. In addition, we can see that 1-3 seconds of audio as an input length to the network is a reasonable choice in the raw waveform model as in the mel-spectrogram model.  

\subsection{Mel-spectrogram and raw waveforms}
Considering that the output size of the first convolution layer in the raw waveform models is equivalent to the mel-spectrogram size, we further validate the effectiveness of the proposed sample-level architecture by performing experiments presented in Table \ref{table:table3}. The models used in the experiments follow the configuration strategy described in Section \ref{sec:modeldesign}. In the mel-spectrogram experiments, 128 mel-bands are used to match up to the number of filters in the first convolution layer of the raw waveform model. FFT size was set to 729 in all comparisons and the magnitude compression is applied with a nonlinear curve, $\log(1+C|A|)$ where $A$ is the magnitude and $C$ is set to 10. 



The results in Table \ref{table:table3} show that the sample-level raw waveform model achieves results comparable to the frame-level mel-spectrogram model. Specifically, we found that using a smaller hop size (81 samples $\approx$ 4 ms) worked better than those of conventional approaches (about 20 ms or so) in the frame-level mel-spectrogram model. However, if the hop size is less than 4 ms, the performance degraded. An interesting finding from the result of the frame-level raw waveform model is that when the filter length is larger than the stride, the accuracy is slightly lower than the models with the same filter length and stride. We interpret that this result is due to the learning ability of the phase variance. As the filter length decreases, the extent of phase variance that the filters should learn is reduced.

\subsection{MSD result and the number of filters}

\begin{figure*}[t]
\centering
\includegraphics[width=\textwidth]{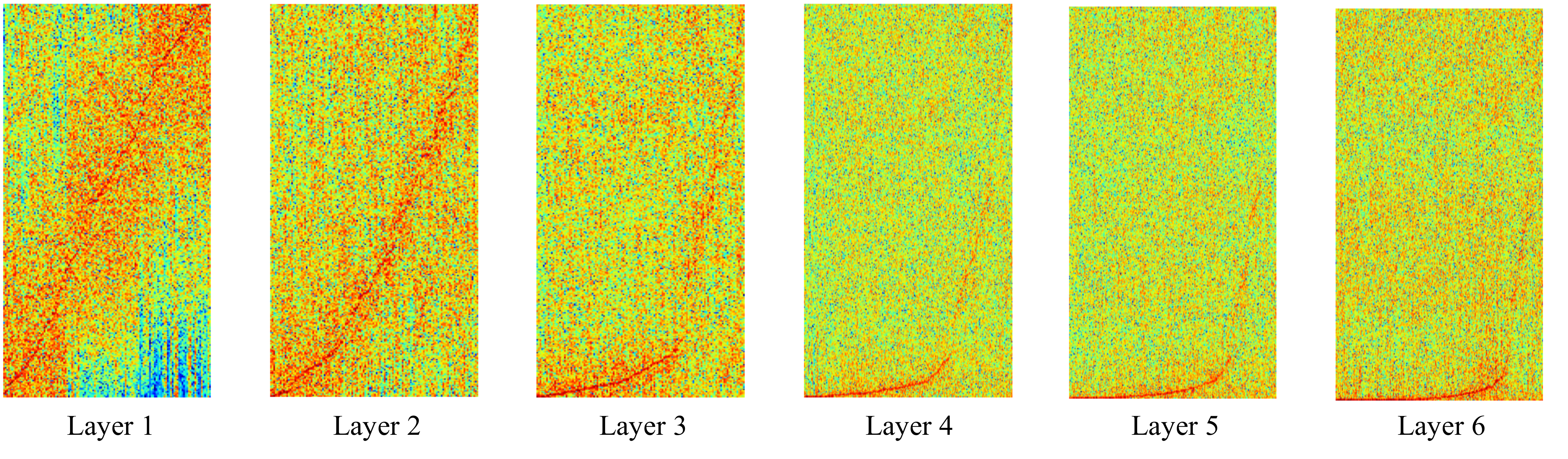}
\caption{Spectrum of the filters in the sample-level convolution layers which are sorted by the frequency of the peak magnitude. The x-axis represents the index of the filter, and the y-axis represents the frequency. The model used for visualization is $3^9$-DCNN with 59049 samples as input. Visualization was performed using the gradient ascent method to obtain the input waveform that maximizes the activation of a filter in the layers. To effectively find the filter characteristics, we set the input waveform estimate to 729 samples which is close to a typical frame size. 
\label{fig:fig2}}
\end{figure*}

We investigate the capacity of our sample-level architecture even further by evaluating the performance on MSD that is ten times larger than MTAT. The result is shown in Table \ref{table:table4}. While training the network on MSD, the number of filters in the convolution layers has been shown to affect the performance. According to our preliminary test results, increasing the number of filters from 16 to 512 along the layers was sufficient for MTAT. However, the test on MSD shows that increasing the number of filters in the first convolution layer improves the performance. Therefore, we increased the number of filters in the first convolution layer from 16 to 128. 

\subsection{Comparison to state-of-the-arts}

In Table \ref{table:table4}, we show the performance of the proposed architecture to previous state-of-the-arts on MTAT and MSD. They show that our proposed sample-level architecture is highly effective compared to them. 

\subsection{Visualization of learned filters}

The technique of visualizing the filters learned at each layer allows better understanding of representation learning in the hierarchical network. However, many previous works in music domain are limited to visualizing learned filters only on the first convolution layer\cite{dieleman2014end,ardilaaudio}. 

\begin{figure}[t]
\centering
\makebox[\columnwidth][c]{\includegraphics[width=1.05\columnwidth]{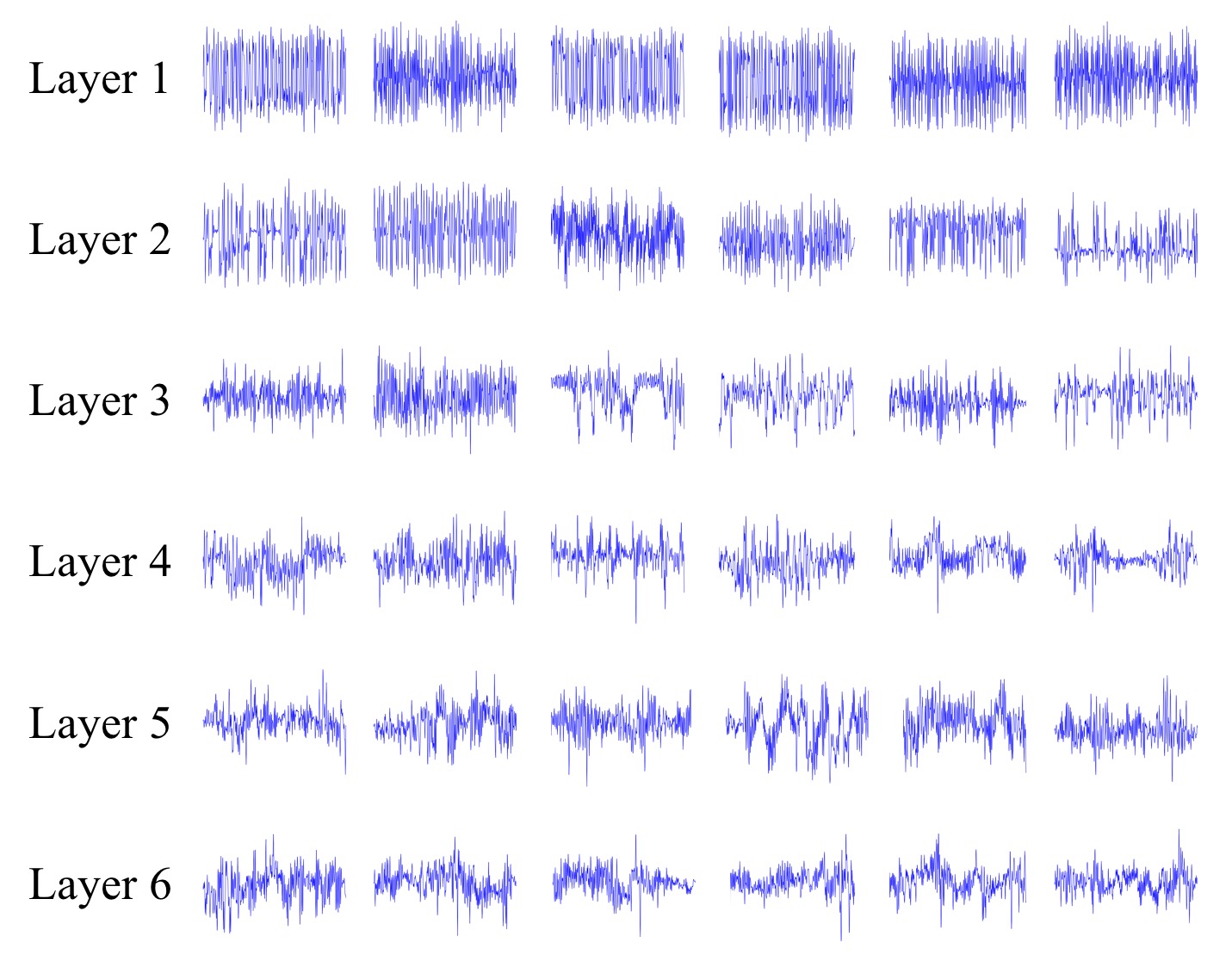}}
\caption{Examples of learned filters at each layer.
\label{fig:fig3}}
\end{figure}

The gradient ascent method has been proposed for filter visualization \cite{erhan2009visualizing} and this technique has provided deeper understanding of what convolutional neural networks learn from images \cite{zeiler2014visualizing,nguyen2015deep}. We applied the technique to our DCNN to observe how each layer hears the raw waveforms. The gradient ascent method is as follows. First, we generate random noise and back-propagate the errors in the network. The loss is set to the target filter activation. Then, we add the bottom gradients to the input with gradient normalization. By repeating this process several times, we can obtain the waveform that maximizes the target filter activation. Examples of learned filters at each layer are in Figure \ref{fig:fig3}. Although we can find the patterns that low-frequency filters are more visible along the layer, estimated filters are still noisy. To show the patterns more clearly, we visualized them as spectrum in the frequency domain and sorted them by the frequency of the peak magnitude.

Note that we set the input waveform estimate to 729 samples in length because, if we initialize and back-propagate to the whole input size of the networks, the estimated filters will have large dimensions such as 59049 samples in computing spectrum. Thus, we used the smaller input samples which can find the filter characteristics more effectively and also are close to a typical frame size in spectrum.   

The layer 1 shows the three distinctive filter bands which are possible with the filter length with 3 samples (say, a DFT size of 3). The center frequency of the filter banks increases linearly in low frequency filter banks but it becomes non-linearly steeper in high frequency filter banks. This trend becomes stronger as the layer goes up. This nonlinearity was found in learned filters with a frame-level end-to-end learning \cite{dieleman2014end} and also in perceptual pitch scales such as mel or bark.    



\section{Conclusion and future work}

In this paper, we proposed sample-level DCNN models that take raw waveforms as input. Through our experiments, we showed that deeper models (more than 10 layers) with a very small sample-level filter length and sub-sampling length are more effective in the music auto-tagging task and the results are comparable to previous state-of-the-art performances on the two datasets. Finally, we visualized hierarchically learned filters. As future work, we will analyze why the deep sample-level architecture works well without input normalization and nonlinear function that compresses the amplitude and also investigate the hierarchically learned filters more thoroughly. 



\begin{acknowledgments}
This work was supported by Korea Advanced Institute of
Science and Technology (project no. G04140049) and National Research Foundation of Korea (project no. N01160463).
\end{acknowledgments} 



\bibliography{smc2017template}

\end{document}